\documentclass[onecolumn,tightenlines,showpacs,nofootinbib,preprintnumbers,
amsmath,amssymb]{revtex4}

\usepackage[usenames,dvipsnames]{color}
\usepackage{caption2}
\usepackage{dcolumn}
\usepackage{graphicx}
\usepackage{bm}
\usepackage{lscape}

\usepackage{setspace}
\begin{document}
\begin{spacing}{1.5}

\title{$CP$ violation induced by the double resonance for pure annihilation decay process in Perturbative QCD}

\author{Gang L\"{u}$^{1}$\footnote{Email: ganglv66@sina.com}, Ye Lu$^{2}$\footnote{Email: luye189@163.com},
Sheng-Tao Li$^{1}$ and Yu-Ting Wang$^{1}$}

\affiliation{\small $^{1}$College of Science, Henan University of Technology, Zhengzhou 450001, China
\\$^{2}$Department of Physics, Guangxi Normal University, Guilin 541004, China
}

\begin{abstract}
In  Perturbative QCD (PQCD) approach we study the direct $CP$ violation in the pure annihilation decay process of
$\bar{B}^0_{s}\rightarrow\pi^+\pi^-\pi^+\pi^-$ induced by the $\rho$ and $\omega$ double resonance effect.
Generally, the $CP$ violation is small in the pure annihilation type decay process. However, we find that the $CP$ violation can be
enhanced by double $\rho-\omega$ interference when the invariant masses of the $\pi^+\pi^-$ pairs are in the vicinity of the $\omega$
resonance. For the decay process of $\bar{B}^0_{s}\rightarrow\pi^+\pi^-\pi^+\pi^-$, the maximum $CP$ violation
can reach 28.64{\%}.\\
\end{abstract}


\maketitle

\section{\label{intro}Introduction}
$CP$ violation is an important area in searching new physics signals beyond the standard model(SM).
It is generally believed that the $B$ meson system provides rich information about $CP$ violation.
The theoretical work has been done in this direction in the past few years.
$CP$ violation arises from the weak phase in the Cabibbo-Kobayasgi-Maskawa (CKM) matrix \cite{cab,kob} in SM.
Meanwhile, it is remarkable that $CP$ violation can still be produced by the interference effects between
the tree and penguin amplitudes. Since the kinematic suppression, the strong phase associated with
long distance rescattering is generally neglected during the past decades.
Recently, the LHCb Collaboration found the large $CP$ violation in the three-body
decay channels of $B^{\pm}\rightarrow \pi^{\pm}\pi^{+}\pi^{-}$
and $B^{\pm}\rightarrow K^{\pm}\pi^{+}\pi^{-}$ \cite{J.M.,R.A.1,R.A.2}.
Hence, the nonleptonic $B$ meson decay from the three-body and four-body decay channels
has been become an important area in searching for $CP$ violation.

A mixing between the $u$ and $d$ flavor leads to the breaking of isospin symmetry for
the $\rho-\omega$ system. The chiral dynamics has been shown restore the isospin symmetry \cite{chi-Nu}.
The $\rho-\omega$ mixing matrix element $\tilde{\Pi}_{\rho\omega}(s)$ gives rise to
isospin violation, where $s$ is the Mandelstam variable.
The magnitude has been extracted by the pion form factor through the cross section of $e^{+}e^{-}\rightarrow \pi^{+}\pi^{-}$.
We can separate the $\tilde{\Pi}_{\rho\omega}(s)$ into two contribution of the
direct coupling of $\omega\rightarrow 2\pi$ and the mixing of $\omega\rightarrow \rho \rightarrow2\pi$.
The emergence of $\tilde{\Pi}_{\rho\omega}(s)$ arises from the inclusion of a nonresonant contribution
to $\omega\rightarrow 2\pi$. The appearance of the $\rho$ and $\omega$ resonance is associated with
complex strong phase from relatively broad $\rho$ resonance region.
Especially, there is perhaps larger strong phase from double $\rho$ and $\omega$ interference.
The $CP$ violation origins from the weak phase difference and the strong phase difference.
Hence, the decay process of $\bar{B}^0_{s}\rightarrow\pi^+\pi^-\pi^+\pi^-$ is a great candidate
for studying the origin of the $CP$ violation.

Meanwhile, it is known that the $CP$ violation
is extremely tiny from the pure annihilation decay process in experiment.
There is relatively large error in dealing with the decay amplitudes from
the QCD factorization approach \cite{qcdf}. The perturbative QCD (PQCD) factorization
approach \cite{Ali:1997nh,9804363,pqcd,pqcd1} is based on $k_{T}$ factorization.
The amplitude can be divided into the convolution of the Wilson coefficients, the light cone wave function,
and hard kernels by the low energy effective Hamiltonian.
The endpoint singularity can be eliminated by introducing the transverse momentum.
However, The transverse momentum integration leads to the double logarithm term which
is resummed into the Sudakov form factor.  The nonperturbative dynamics are included in the meson
wave function which can be extracted from experiment.
The hard one can be calculated by perturbation theory.

The remainder of this paper is organized as follows. In Sec. \ref{sec:hamckm} we present the form of the effective Hamiltonian.
In Sec. \ref{sec:cpv1} we give the calculating formalism and calculation details of $CP$ violation from $\rho-\omega$ mixing
in the $\bar{B}^0_{s}\rightarrow \rho^0(\omega)\rho^0(\omega)\rightarrow\pi^+\pi^-\pi^+\pi^-$ decay. In Sec. \ref{int} we show input parameters.
We present the numerical results in Sec. \ref{num}.
Summary and discussion are included in
Sec. \ref{sec:conclusion}. The related function defined in the text are given
in the Appendix.

\section{\label{sec:hamckm}The effective hamiltonian}

With the operator product expansion, the effective weak Hamiltonian can
be written as \cite{buch}
\begin{eqnarray}
 {\cal H}_{eff} &=& \frac{G_{F}}{\sqrt{2}}
     \Bigg\{ V_{ub} V_{uq}^{\ast} \Big[
     C_{1}({\mu}) Q^{u}_{1}({\mu})
  +  C_{2}({\mu}) Q^{u}_{2}({\mu})\Big]
  -V_{tb} V_{tq}^{\ast} \Big[{\sum\limits_{i=3}^{10}} C_{i}({\mu}) Q_{i}({\mu})
  \Big ] \Bigg\} + \mbox{H.c.} ,
 \label{2a}
 \vspace{2mm}
 \end{eqnarray}
where $q=(d, s)$, $G_F$ represents Fermi constant, $C_i$ (i=1,...,10) are the Wilson coefficients,
$V_{q_1q_2}$ ($q_1$ and $q_2$ represent quarks) is the CKM matrix element, and $O_i$ is the four quark operator. The
operators $O_i$ have the following forms:
\begin{eqnarray}
O^{u}_1&=& \bar d_\alpha \gamma_\mu(1-\gamma_5)u_\beta\bar
u_\beta\gamma^\mu(1-\gamma_5)b_\alpha,\nonumber\\
O^{u}_2&=& \bar d \gamma_\mu(1-\gamma_5)u\bar
u\gamma^\mu(1-\gamma_5)b,\nonumber\\
O_3&=& \bar d \gamma_\mu(1-\gamma_5)b \sum_{q'}
\bar q' \gamma^\mu(1-\gamma_5) q',\nonumber\\
O_4 &=& \bar d_\alpha \gamma_\mu(1-\gamma_5)b_\beta \sum_{q'}
\bar q'_\beta \gamma^\mu(1-\gamma_5) q'_\alpha,\nonumber\\
O_5&=&\bar d \gamma_\mu(1-\gamma_5)b \sum_{q'} \bar q'
\gamma^\mu(1+\gamma_5)q',\nonumber\\
O_6& = &\bar d_\alpha \gamma_\mu(1-\gamma_5)b_\beta \sum_{q'}
\bar q'_\beta \gamma^\mu(1+\gamma_5) q'_\alpha,\nonumber\\
O_7&=& \frac{3}{2}\bar d \gamma_\mu(1-\gamma_5)b \sum_{q'}
e_{q'}\bar q' \gamma^\mu(1+\gamma_5) q',\nonumber\\
O_8 &=&\frac{3}{2} \bar d_\alpha \gamma_\mu(1-\gamma_5)b_\beta \sum_{q'}
e_{q'}\bar q'_\beta \gamma^\mu(1+\gamma_5) q'_\alpha,\nonumber\\
O_9&=&\frac{3}{2}\bar d \gamma_\mu(1-\gamma_5)b \sum_{q'} e_{q'}\bar q'
\gamma^\mu(1-\gamma_5)q',\nonumber\\
O_{10}& = &\frac{3}{2}\bar d_\alpha \gamma_\mu(1-\gamma_5)b_\beta \sum_{q'}
e_{q'}\bar q'_\beta \gamma^\mu(1-\gamma_5) q'_\alpha,
\label{2b}
\vspace{2mm}
\end{eqnarray}
where $\alpha$ and $\beta$ are color indices, and $q^\prime=u, d, s, c$
or $b$ quarks. In Eq.(\ref{2b}) $O_1^u$ and $O_2^u$ are tree
operators, $O_3$--$O_6$ are QCD penguin operators and $O_7$--$O_{10}$ are
the operators associated with electroweak penguin diagrams. $C_i(m_b)$ can be written \cite{pqcd1},
\begin{eqnarray}
C_1 &=&-0.2703, \;\; \;C_2=1.1188,\nonumber\\
C_3 &=& 0.0126,\;\;\;C_4 = -0.0270,\nonumber\\
C_5 &=& 0.0085,\;\;\;C_6 = -0.0326,\nonumber\\
C_7 &=& 0.0011,\;\;\;C_8 = 0.0004,\nonumber\\
C_9&=& -0.0090,\;\;\;C_{10} = 0.0022.
\label{2k}
\vspace{2mm}
\end{eqnarray}
So, we can obtain numerical values of $a_i$.
The combinations $a_i$ of Wilson coefficients are
defined as usual \cite{9804363}:
\begin{eqnarray}
a_1&=&C_2+C_1/3,\;\; \;a_2=C_1+C_2/3,\nonumber \\
a_3&=&C_3+C_4/3,\;\; \;a_4=C_4+C_3/3,\nonumber \\
a_5&=&C_5+C_6/3,\;\; \;a_6=C_6+C_5/3,\nonumber \\
a_7&=&C_7+C_8/3,\;\; \;a_8=C_8+C_7/3,\nonumber \\
a_9&=&C_9+C_{10}/3,\;\; \;a_{10}= C_{10}+C_{9}/3.
\label{2k}
\vspace{2mm}
\end{eqnarray}

\section{\label{sec:cpv1}$CP$ violation in $\bar{B}_{s}^{0}\rightarrow \rho^0(\omega)\rho^0(\omega)\rightarrow \pi^+\pi^{-}\pi^+\pi^{-}$}
\subsection{\label{subsec:form}Formalism}

The amplitudes $A^{\sigma}$ of the process $\bar B_s(p) \to V_1(p_1,\epsilon_{1}) +
V_2(p_2,\epsilon_{2})$ can be written {\cite{Kram1991}}
\begin{equation}
A^{\sigma}=\epsilon_{1\mu}^{*}({\sigma})\epsilon_{2\nu}^{*}({\sigma})(ag^{\mu\nu}+\frac{b}{m_{1}m_{2}}p^{\mu}p^{\nu}+\frac{ic}{m_{1}m_{2}}\epsilon^{\mu\nu\alpha\beta}p_{1\alpha}p_{2\beta})
\end{equation}
where $\sigma$ is the helicity of the vector meson. $\epsilon_1$($p_1$) and $\epsilon_2$($p_2$) are the
polarization vectors (momenta) of $V_{1}$ and $V_{2}$, respectively.
$m_{1}$ and $m_{2}$ refer to the masses of the vector mesons $V_{1}$ and $V_{2}$.
The invariant amplitudes a, b, c are associated with the amplitude $A_{i}$ (
i refer to the three kind of polarizations, longitudinal (L), normal (N) and transverse (T)).
Then we have
\begin{equation}
A^{\sigma}=M^2_{B_{s}}A_{L}+M^2_{B_{s}}A_{N}\epsilon_{1\mu}^{*}(\sigma=T)\cdot \epsilon_{2\mu}^{*}(\sigma=T)+iA_{T}
\epsilon^{\alpha\beta\gamma\rho}\epsilon_{1\alpha}^{*}(\sigma)\epsilon_{2\alpha}^{*}(\sigma)p_{1\gamma}p_{2\rho}
\end{equation}
The longitudinal $H_{0}$, transverse $H_{\pm}$ of helicity amplitudes can be expressed
$H_{0}=M^2_{B_{s}}A_{L}$,
$H_{\pm}=M^2_{B_{s}}A_{N}\mp m_{1}m_{2} \sqrt{r^2-1}A_{T}$.
The decay width is written
\begin{equation}
\Gamma =\frac{P_c}{8\pi M^{2}_{B_s} }
A^{(\sigma)+}A^{(\sigma)}=\frac{P_c}{8\pi M^{2}_{B_s} }|H_{0}|^{2}+|H_{+}|^{2}+|H_{-}|^{2}. \label{dr1}
\end{equation}

The interaction of the photon and the hadronic matter can be described by the vector meson dominance model (VMD)
{\cite{Sakurai1969}}. The photon can couple to the hadronic field through a $\rho$ meson.
The mixing matrix element $\widetilde{\Pi}_{\rho\omega}(s)$ is extracted from the data of the
cross section for $e^{+}e^{-}\rightarrow \pi^{+}\pi^{-}$ {\cite{Connell1997,Connell1997-1}}.
The nonresonant contribution of $\omega\rightarrow\pi^+\pi^-$ has been
effectively absorbed into $\widetilde{\Pi}_{\rho\omega}$
 which leads to the explicit $s$ dependence of
$\widetilde{\Pi}_{\rho\omega}$ \cite{oco}.
We can make the expansion
$\widetilde{\Pi}_{\rho\omega}(s)=\widetilde{\Pi}_{\rho\omega}(m_{\omega}^2)+(s-m_{\omega})\widetilde{\Pi}_{\rho\omega}^\prime(m_{\omega}^2)$.
However, one can neglect the $s$ dependence of $\widetilde{\Pi}_{\rho\omega}$  in practice.
The $\rho-\omega$ mixing parameters were determined in the
fit of Gardner and O'Connell \cite{gard}:
\begin{eqnarray}
\mathfrak{Re}\widetilde{\Pi}_{\rho\omega}(m_{\omega}^2)&=&-3500\pm300
\rm{MeV}^2,\nonumber\\{\mathfrak{Im}}\widetilde{\Pi}_{\rho\omega}(m_{\omega}^2)&=&-300\pm300
\textrm{MeV}^2,\nonumber\\\widetilde{\Pi}_{\rho\omega}^\prime(m_{\omega}^2)&=&0.03\pm0.04.
\end{eqnarray}

The formalism of the $CP$ violation is presented for the $\bar{B}_{s}^{0}$ meson decay process in the
following. The amplitude $A$ ($\bar{A}$) for the decay process
$\bar{B}_{s}^{0}\rightarrow\pi^+\pi^{-}\pi^+\pi^{-}$
(${B}_{s}^{0}\rightarrow\pi^+\pi^{-}\pi^+\pi^{-}$) can be written as:
\begin{eqnarray}
A=<\pi^+\pi^{-}\pi^+\pi^{-}|H^T|\bar{B}_{s}^{0}>+<\pi^+\pi^{-}\pi^+\pi^{-}|H^P|\bar{B}_{s}^{0}>,\label{A}
\end{eqnarray}
\begin{eqnarray}
\bar{A}=<\pi^+\pi^{-}\pi^+\pi^{-}|H^T|{B}_{s}^{0}>+<\pi^+\pi^-\pi^+\pi^{-}|H^P|{B}_{s}^{0}>,\label{Abar}
\end{eqnarray}
where $H^T$ and $H^P$ refer to the tree and
penguin operators in the Hamiltonian, respectively.
We define the relative magnitudes and phases between the tree
and penguin operator contributions as follows:
\begin{eqnarray}
A=\big<\pi^+\pi^{-}\pi^+\pi^{-}|H^T|\bar{B}_{s}^{0}\big>[1+re^{i(\delta+\phi)}],\label{Abar}\\
\bar{A}=\big<\pi^+\pi^-\pi^+\pi^{-}|H^T|{B}_{s}^{0}\big>[1+re^{i(\delta-\phi)}],
\label{A'bar}
\end{eqnarray}
where $\delta$ and $\phi$ are strong and weak phases, respectively.
The weak phase difference $\phi$ can be expressed as a combination of the CKM matrix elements:
$\phi=\arg[(V_{tb}V_{ts}^{*})/(V_{ub}V_{us}^{*})]$. The
parameter $r$ is the absolute value of the ratio of tree and penguin
amplitudes:
\begin{eqnarray}
r\equiv\Bigg|\frac{\big<\pi^+\pi^{-}\pi^+\pi^{-}|H^P|\bar{B}_{s}^{0}\big>}{\big<\pi^+\pi^{-}\pi^+\pi^{-}|H^T|\bar{B}_{s}^{0}\big>}\Bigg|
\label{r}.
\end{eqnarray}
The parameter of $CP$ violating asymmetry, $A_{cp}$, can be written as
\begin{eqnarray}
A_{CP}=\frac{|A|^{2}-|\bar{A}|^{2}}{|A|^{2}+|\bar{A}|^{2}}
  =\frac{-2(T_{0}^2r_{0}\sin\delta_0+T_{+}^2r_{+}\sin\delta_{+}+T_{-}^2r_{-}\sin\delta_-)\sin\phi}
  {\sum_{i=0+-}T_{i}^2(1+r_{i}^2+2r_{i}\cos\delta_i\cos\phi)},
\label{eq:CP-tuidao}
\end{eqnarray}
where
\begin{eqnarray}
|A|^{2}=\sum_{\sigma}A^{(\sigma)+}A^{(\sigma)}=|H_{0}|^{2}+|H_{+}|^{2}+|H_{-}|^{2}
\end{eqnarray}
and $T_{i}(i=0,+,-)$ represent the tree-level helicity amplitudes.
We can see explicitly from Eq. (\ref{eq:CP-tuidao}) that both weak and strong phase
differences are responsible for $CP$ violation.
$\rho-\omega$ mixing introduces the strong phase difference and well known in the three body decay processes of the bottom hadron
\cite{guo1,guo11,lei,guo2,gang1,gang2,gang3}. Due to $\rho-\omega$ interference from the u and d quark mixing, we can
write the following formalism in an approximate from the first order of isospin violation:
\begin{eqnarray}
\big<\pi^+\pi^-\pi^+\pi^-|H^T|\bar{B}^0_{s}\big>=\frac{2g_{\rho}^2}{s_{\rho}^{2}s_{\omega}}\widetilde{\Pi}_{\rho\omega}t_{\rho\omega}+\frac{g_{\rho}^2}{s_{\rho}^2}t_{\rho\rho},
\label{Htr}\\
\big<\pi^+\pi^-\pi^+\pi^-|H^P|\bar{B}^0_{s}\big>=\frac{2g_{\rho}^2}{s_{\rho}^{2}s_{\omega}}\widetilde{\Pi}_{\rho\omega}p_{\rho\omega}+\frac{g_{\rho}^2}{s_{\rho}^2}p_{\rho\rho},
\label{Hpe}
\end{eqnarray}
where $t_{\rho\rho}(p_{\rho\rho})$ and
$t_{\rho\omega}(p_{\rho\omega})$ are the tree (penguin) amplitudes
for $\bar{B}_{s}\rightarrow\rho^0\rho^0$ and
$\bar{B}_{s}\rightarrow\rho^0\omega$, respectively, $g_{\rho}$ is
the coupling for $\rho^0\rightarrow\pi^+\pi^-$,
$\widetilde{\Pi}_{\rho\omega}$ is the effective $\rho-\omega$
mixing amplitude which also effectively includes the direct
coupling $\omega\rightarrow\pi^+\pi^-$. $s_{V}$, $m_{V}$ and $\Gamma_V$($V$=$\rho$ or
$\omega$) is the inverse propagator, mass and decay rate of the vector meson $V$, respectively.
\begin{eqnarray}
s_V=s-m_V^2+{\rm{i}}m_V\Gamma_V,
\end{eqnarray}
with $\sqrt{s}$ being the invariant masses of the $\pi^+\pi^-$
pairs. There are double $\rho- \omega$ interference in the decay process of $\bar{B}^0_{s}\rightarrow \rho^0(\omega)\rho^0(\omega)\rightarrow\pi^+\pi^-\pi^+\pi^-$.
Hence, a factor of 2 appears in Eqs. (\ref{Htr}), (\ref{Hpe}) compared with the case of single $\rho- \omega$
interference \cite{eno,gar,guo1,guo2,guo11,lei,gang1,gang2,gang3}.
From Eqs. (\ref{A})(\ref{Abar})(\ref{Htr})(\ref{Hpe}) one has
\begin{eqnarray}
re^{i\delta}e^{i\phi}=\frac{2\widetilde{\Pi}_{\rho\omega}p_{\rho\omega}+s_{\omega}p_{\rho\rho}}{2\widetilde{\Pi}_{\rho\omega}t_{\rho\omega}+s_{\omega}t_{\rho\rho}},
\label{rdtdirive}
\end{eqnarray}
Defining
\begin{eqnarray}
\frac{p_{\rho\omega}}{t_{\rho\rho}}\equiv r^\prime
e^{i(\delta_q+\phi)},\quad\frac{t_{\rho\omega}}{t_{\rho\rho}}\equiv
\alpha
e^{i\delta_\alpha},\quad\frac{p_{\rho\rho}}{p_{\rho\omega}}\equiv
\beta e^{i\delta_\beta}, \label{def}
\end{eqnarray}
where $\delta_\alpha$, $\delta_\beta$ and $\delta_q$ are strong
phases, one finds the following expression from Eqs.
(\ref{rdtdirive})(\ref{def}):
\begin{eqnarray}
re^{i\delta}=r^\prime
e^{i\delta_q}\frac{2\widetilde{\Pi}_{\rho\omega}+\beta
e^{i\delta_\beta}s_{\omega}}{2\widetilde{\Pi}_{\rho\omega}\alpha
e^{i\delta_\alpha}+s_{\omega}}. \label{rdt}
\end{eqnarray}
In order to obtain the $CP$ violating asymmetry in Eq.
(\ref{eq:CP-tuidao}), sin$\phi$ and cos$\phi$ are needed, where $\phi$ is
determined by the CKM matrix elements. In the Wolfenstein
parametrization \cite{wol}, one has
\begin{eqnarray}
{\rm sin}\phi &=&-\frac{\eta }{\sqrt{\rho ^2+\eta ^2}}, \nonumber \\
{\rm cos}\phi &=&-\frac{\rho }{\sqrt{\rho ^2+\eta ^2}}.
\label{3l1}
\vspace{2mm}
\end{eqnarray}

\subsection{\label{cal}Calculation details}

We can decompose the decay amplitude for the decay process $\bar{B}_{s}^0\rightarrow \rho^{0}(\omega)\rho^{0}(\omega)$
in terms of tree-level and penguin-level contributions depending on
the CKM matrix elements of $V_{ub}V^{*}_{us}$ and $V_{tb}V^{*}_{ts}$.
Due to the equations (\ref{eq:CP-tuidao})(\ref{rdtdirive})(\ref{def}), we calculate the
amplitudes $t_{\rho\rho}$, $t_{\rho\omega}$, $p_{\rho\rho}$ and $p_{\rho\omega}$ in perturbative QCD approach.
The $F$ and $M$ function associated with the decay amplitudes can be found in the appendix from the perturbative QCD approach.

There are four types of Feynman diagrams contributing to $\bar{B}_{s} \to M_{2}M_{3}$($M_{2}$,$M_{3}$=$\rho$ or $\omega$) annihilation decay mode at leading order.
The pure annihilation type process can be classified into factorizable diagrams and
non-factorizable diagrams \cite{kphi,krho}. Through calculating these diagrams, we can
get the amplitudes $A^{(i)}$, where $i=L, N, T$ standing for the longitudinal and two transverse polarizations. Because these diagrams are the same as those of $B\to K^*\phi$ and $B\to K^*\rho$ decays \cite{kphi,krho}, the formulas of $\bar{B}_{s} \to \rho \rho$ or $\bar{B}_{s} \to \rho \omega$ are similar to those of $B\to K^*\phi$ and $B\to K^*\rho$. We just need to replace some corresponding wave functions, Wilson coefficients and corresponding parameters. 

With the Hamiltonian (\ref{2a}), depending on CKM matrix elements of $V_{ub}V^{*}_{us}$ and  $V_{tb}V^{*}_{ts}$,
the decay amplitudes $A^{(i)}(i=L, N, T)$ for $\bar B_{s}^0\rightarrow \rho^{0}\rho^{0}$ in PQCD can be
written as
\begin{eqnarray}
\sqrt{2}A^{(i)}(\bar B_{s}^0\to\rho^{0}\rho^{0})=V_{ub}V_{us}^{*}t_{\rho\rho}^{i}-V_{tb}V_{ts}^{*}p_{\rho\rho}^{i}, \label{BcDrho1}
\end{eqnarray}
The tree level amplitude $t_{\rho\rho}$ can written as
\begin{eqnarray}
t_{\rho\rho}^{i}&=&\frac{G_F}{\sqrt{2}} \Big \{ f_{B_s}
F_{ann}^{LL,i}\left[a_{2}\right]
+ M_{ann}^{LL,i}[C_2]\Big\},  \label{trho1}
\end{eqnarray}
where $f_{B_s}$ refers to the decay constant of $\bar B_{s}$ meson.

The penguin level amplitude are expressed in the following
\begin{eqnarray}
p_{\rho\rho}^{i}&=&\frac{G_F}{\sqrt{2}} \bigg \{ f_{B_s}
F_{ann}^{LL,i}\left[2a_{3} +\frac{1}{2}a_{9} \right] +f_{B_s}
F_{ann}^{LR,i}\left[2a_{5} +\frac{1}{2}a_{7} \right]
  \nonumber\\
&&+M_{ann}^{LL,i}\left[2C_{4}+\frac{1}{2}C_{10}\right]
+M_{ann}^{SP,i}\left[2C_{6}+\frac{1}{2}C_{8}\right]\bigg \}.  \label{Prho1}
\end{eqnarray}
The decay amplitude for $\bar B_{s}^0\to \rho^{0}\omega$ can be written as
\begin{eqnarray}
2A^{(i)}(\bar B_{s}^0\to\rho^{0}\omega)&=& V_{ub}V_{us}^{*} t_{\rho\omega}^{i} -  V_{tb}V_{ts}^{*} p_{\rho\omega}^{i}. \label{BcDomega1}
\end{eqnarray}
We can give the tree level the contribution in the following
\begin{eqnarray}
t_{\rho\omega}^{i}&=&\frac{G_F}{\sqrt{2}}\Big\{ f_{B_s}F_{ann}^{LL,i}\left[a_{2}\right] +  M_{ann}^{LL,i}[C_{2}] \Big\}, \label{tomega1}
\end{eqnarray}
and the penguin level contribution are given as following
\begin{eqnarray}
p_{\rho\omega}^{i}&=& \frac{G_F}{\sqrt{2}}  V_{tb}V_{ts}^{*}\Big\{ f_{B_s}
F_{ann}^{LL,i}\left[\frac{3}{2}a_9 \right]+ f_{B_s}F_{ann}^{LR,i}\left[\frac{3}{2}a_7 \right]  \nonumber\\
&&+M_{ann}^{LL,i}\left[\frac{3}{2}C_{10}\right]
+M_{ann}^{SP,i}\left[\frac{3}{2}C_8\right]\Big\} +\left [\rho^0
\leftrightarrow \omega\right ]. \label{pomega1}
\end{eqnarray}
Based on the definition of (\ref{def}), we can get
\begin{eqnarray}
\alpha e^{i\delta_\alpha}&=&\frac{t_{\rho\omega}}{t_{\rho\rho}}, \label{eq:afaform} \\
\beta e^{i\delta_\beta}&=&\frac{p_{\rho\rho}}{p_{\rho\omega}}, \label{eq:btaform}\\
r^\prime e^{i\delta_q}&=&\frac{p_{\rho\omega}}{t_{\rho\rho}}
\times\bigg|\frac{V_{tb}V_{ts}^*}{V_{ub}V_{us}^*}\bigg|,  \label{eq:delform}
\end{eqnarray}
where
\begin{equation}
\left|\frac{V_{tb}V^{*}_{ts}}{V_{ub}V^{*}_{us}}\right|=\frac{\sqrt{\rho^2+\eta^2}}{\lambda^2(\rho^2+\eta^2)}.
\label{3p}
\vspace{2mm}
\end{equation}

\section{\label{int}Input parameters}
The CKM matrix, which elements are determined from experiments, can be expressed in terms of the Wolfenstein parameters $A$, $\rho$, $\lambda$ and $\eta$ \cite{wol}:
\begin{equation}
\left(
\begin{array}{ccc}
  1-\tfrac{1}{2}\lambda^2   & \lambda                  &A\lambda^3(\rho-\mathrm{i}\eta) \\
  -\lambda                 & 1-\tfrac{1}{2}\lambda^2   &A\lambda^2 \\
  A\lambda^3(1-\rho-\mathrm{i}\eta) & -A\lambda^2              &1\\
\end{array}
\right),\label{ckm}
\end{equation}
where $\mathcal{O} (\lambda^{4})$ corrections are neglected. The latest values for the parameters in the CKM matrix are \cite{ganglvpqcdbc}:
\begin{eqnarray}
&& \lambda=0.22537\pm0.00061,\quad A=0.814_{-0.024}^{+0.023},\nonumber \\
&& \bar{\rho}=0.117\pm0.21,\quad
\bar{\eta}=0.353\pm{+0.013}.\label{eq: rhobarvalue}
\end{eqnarray}
where
\begin{eqnarray}
 \bar{\rho}=\rho(1-\frac{\lambda^2}{2}),\quad
\bar{\eta}=\eta(1-\frac{\lambda^2}{2}).\label{eq: rho rhobar
relation}
\end{eqnarray}
From Eqs. (\ref{eq: rhobarvalue}) ( \ref{eq: rho rhobar relation})
we have
\begin{eqnarray}
0.121<\rho<0.158,\quad  0.336<\eta<0.363.\label{eq: rho value}
\end{eqnarray}
%
The other parameters and the corresponding references are listed in Table.1.
\begin{table}[h]
\caption{\label{InputParameters}Input parameters used in this paper.}
\begin{tabular}{|*{8}{l|}}
\hline
Parameters&Input data & References  \\ \hline
Fermi constant (in $\text{GeV}^{-2}$)&$G_F=1.16638\times10^{-5}$& \cite{PDG2016}\\ \hline
                        &$m_{B^0_s}=5.36677,~\tau_{B^0_s}=1.512\times10^{-12}s$& \\
                        &$m_{\rho^0(770)}=0.77526, ~\Gamma_{\rho^0(770)}=0.1491,$&\\
Masses and decay widths&$m_{\omega(782)}=0.78265, ~\Gamma_{\omega(782)}=8.49\times10^{-3},$& \cite{PDG2016}\\
(in GeV)                &$m_\pi=0.13957,~m_W=80.385,$&\\
                        &$m_u=0.0023,~m_d=0.0048,$&\\
                        &$m_s=0.095,~m_c=1.275,$&\\
                        &$m_t=173.21,~m_b=4.18,$&\\\hline
Decay constants&$f_\rho=209\pm2,~f_\rho^T=165\pm9,$&\cite{PDG2016,Li:2006jv}\\
(in MeV)       &$f_\omega=195.1\pm3,~f_\omega^T=145\pm10,$& \\ \hline
\end{tabular}\label{table1}
\end{table}
\section{\label{num}The numerical results of $CP$ violation in $\bar{B}^0_{s}\rightarrow \rho^0(\omega)\rho^0(\omega)\rightarrow\pi^+\pi^-\pi^+\pi^-$}

\begin{figure}[h]
\includegraphics[width=0.5\textwidth]{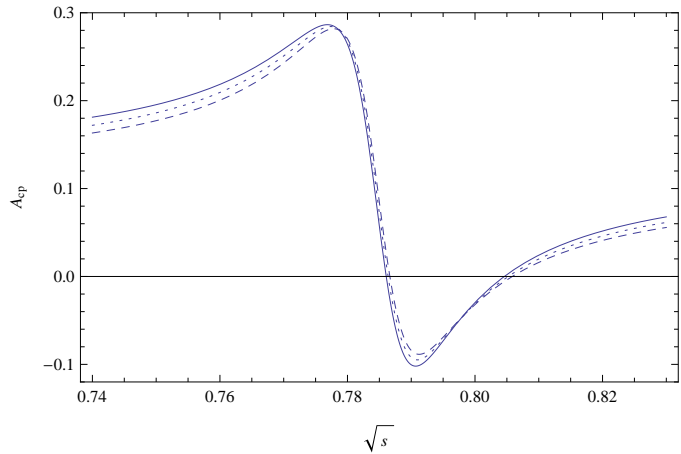}\\
\caption{The $CP$ violating asymmetry, $A_{cp}$, as a function of $\sqrt{s}$ for
different CKM matrix elements. The dash line, dot line and solid line corresponds to the maximum, middle, and minimum CKM matrix element for the decay channel of $\bar{B}^0_{s}\rightarrow \rho^0(\omega)\rho^0(\omega)\rightarrow\pi^+\pi^-\pi^+\pi^-$, respectively.} \label{Acp plot}
\end{figure}

\begin{figure}
\includegraphics[width=0.5\textwidth]{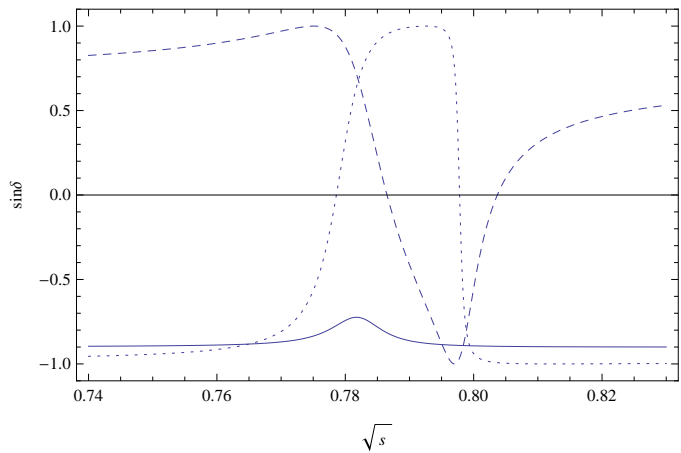}\\
\caption{$sin\delta$ as a function of $\sqrt{s}$
corresponding to central parameter values of CKM matrix elements
for $\bar{B}^0_{s}\rightarrow \rho^0(\omega)\rho^0(\omega)\rightarrow\pi^+\pi^-\pi^+\pi^-$.
The dash line, dot line and solid line corresponds to $sin{\delta_{0}}$, $sin{\delta_{+}}$ and $sin{\delta_{-}}$, respectively.}\label{sin1 plot}
\end{figure}

\begin{figure}
\includegraphics[width=0.5\textwidth]{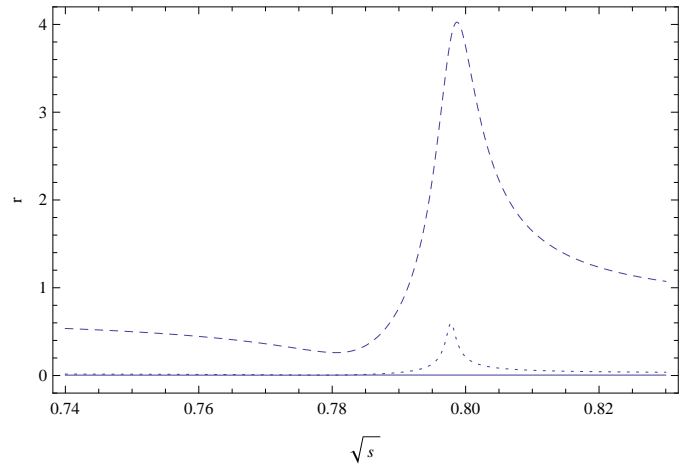}\\
\caption{Plot of $r$ as a function of $\sqrt{s}$
corresponding to central parameter values of CKM matrix elements
for $\bar{B}^0_{s}\rightarrow \rho^0(\omega)\rho^0(\omega)\rightarrow\pi^+\pi^-\pi^+\pi^-$.
The dash line, dot line and solid line corresponds to $r_{0}$, $r_{+}$ and $r_{-}$, respectively.}\label{r plot}
\end{figure}

In the numerical results, we find that the $CP$ violation can be enhanced via double $\rho-\omega$ mixing for the pure annihilation type decay channel $\bar{B}^0_{s}\rightarrow \rho^0(\omega)\rho^0(\omega)\rightarrow\pi^+\pi^-\pi^+\pi^-$ when the invariant mass of $\pi^{+}\pi^{-}$ is in the vicinity of the
$\omega$ resonance within perturbative QCD scheme. The $CP$ violation depends on the weak phase difference from CKM matrix elements and the strong phase difference which is difficult to control. The CKM matrix elements, which relate to $\rho$, $\eta$, $\lambda$ and $A$, are given in Eq.(\ref{eq: rhobarvalue}). The uncertainties due to the CKM matrix elements come from $\rho$, $\eta$, $\lambda$ and $A$. In our numerical calculations, we let $\rho$, $\eta$, $\lambda$ and $A$ vary among the limiting values.
The numerical results are shown from Fig.~\ref{Acp plot} to Fig.~\ref{r plot} with the different parameter values of CKM matrix elements. The dash line, dot line and solid line corresponds to the maximum, middle, and minimum CKM matrix element for the decay channel of $\bar{B}^0_{s}\rightarrow \rho^0(\omega)\rho^0(\omega)\rightarrow\pi^+\pi^-\pi^+\pi^-$, respectively. We find the results are not sensitive to the  values of $\rho$, $\eta$, $\lambda$ and $A$. In Fig.~\ref{Acp plot}, we give the plot of $CP$ violating asymmetry as a function of $\sqrt{s}$. From the Fig.~\ref{Acp plot}, one can see the $CP$ violation parameter is dependent on $\sqrt{s}$ and changes rapidly due to $\rho-\omega$ mixing when the invariant mass of $\pi^{+}\pi^{-}$ is in the vicinity of the $\omega$ resonance. From the numerical results, it is found that the maximum $CP$ violating parameter reaches $28.64\%$ in the case of ($\rho_{mini}$, $\eta_{mini}$).

From Eq.(\ref{eq:CP-tuidao}), one can see that the $CP$ violating parameter depend on both sin$\delta$ and $r$. The plots of
$\sin\delta$ and $r$ as a function of $\sqrt{s}$ are shown in Fig.~\ref{sin1 plot}, and Fig.~\ref{r plot}, respectively. It
can be seen that $\sin\delta_{0}$ ($sin\delta_{-}$ and $sin\delta_{+}$) vary sharply at the range of the resonance in Fig.~\ref{sin1 plot}.
One can see that $r$ change largely in the vicinity of the $\omega$ resonance.

\section{\label{sec:conclusion}Summary and conclusion}

In this paper, we study the $CP$ violation for the pure annihilation type decay process of $\bar{B}^0_{s}\rightarrow \pi^+\pi^-\pi^+\pi^-$ in perturbative QCD.
It has been found that the $CP$ violation can be enhanced greatly at the area of $\rho-\omega$ resonance.
The maximum $CP$ violation value can reach $28.64\%$ due to double $\rho$ and $\omega$ resonance.

The theoretical errors are large which follows to
the uncertainties of results. Generally, power corrections beyond the heavy quark limit give the major theoretical uncertainties. This implies the necessity of introducing $1/m_b$ power corrections. Unfortunately, there are many possible $1/m_b$ power suppressed effects and they are generally nonperturbative in nature and hence not calculable by the perturbative method. There are more uncertainties in this scheme. The first error refers to the variation of the CKM parameters, which are given in Eq.(\ref{eq: rhobarvalue}). The second error comes from the hadronic parameters: the shape parameters, form factors, decay constants, and the wave function of the $B_s$ meson. The third error corresponds to the choice of the hard scales, which vary from 0.75t to 1.25t, which characterizing the size of next-to-leading order QCD contributions. Therefore, the results for $CP$ violating asymmetrie of the 
decay process $\bar{B}^0_{s}\rightarrow \pi^+\pi^-\pi^+\pi^-$  is given as following:
\begin{eqnarray}
A_{CP}(\bar{B}^0_{s}\rightarrow \pi^+\pi^-\pi^+\pi^-)=28.43^{+0.21+0.25+5.62}_{-0.25-0.16-3.98}\%,
\end{eqnarray}
where the first uncertainty is corresponding to the CKM parameters, the second comes from the hadronic parameters, and the third is associated with the hard scales.
The LHC experiment may detect the large $CP$ violation for the decay process $\bar{B}^0_{s}\rightarrow \pi^+\pi^-\pi^+\pi^-$ in the region of the $\omega$ resonance.

\section{APPENDIX: Related functions defined in the text}

In this appendix we present explicit expressions of the factorizable and non-factorizable amplitudes with Perturbative QCD in  Eq.(\ref{BcDrho1}) and Eq.(\ref{BcDomega1}) \cite{pqcd,pqcd1,Lcd:the two-body,prd81014022}. The factorizable amplitudes $F_{ann}^{LL,i}(a_i)$, and $F_{ann}^{SP,i}(a_i)$ (i=L,N,T) are written as

\begin{eqnarray}
f_{B_s}F_{ann}^{LL,N}(a_i)&=& f_{B_s}F_{ann}^{LR,N}(a_i)
\end{eqnarray}

\begin{eqnarray}
f_{B_s}F_{ann}^{LL,N}(a_i)
&=& -8\pi C_FM_{B_s}^4f_{B_s}r_2
r_3\int^1_0dx_2dx_3\int^\infty_0b_2db_2b_3db_3
\Big\{E_a(t_c)a_i(t_c)h_a(x_2,1-x_3,b_2,b_3)
)\nonumber \\
&&\left[(2-x_3)\left(\phi_2^v(x_2)\phi_3^v(x_3)+\phi_2^a(x_2)\phi_3^a(x_3)\right)
+x_3(\phi_2^v(x_2)\phi_3^a(x_3)+\phi_2^a(x_2)\phi_3^v(x_3))\right]
\nonumber \\
&&-h_a(1-x_3,x_2,b_3,b_2)[(1+x_2)
(\phi_2^v(x_2)\phi_3^v(x_3)+\phi_2^a(x_2)\phi_3^a(x_3))
\nonumber \\
&&
-(1-x_2)(\phi_2^v(x_2)\phi_3^a(x_3)+\phi_2^a(x_2)\phi_3^v(x_3))]
E_a(t_c')a_i(t_c')\Big\}.
\end{eqnarray}

\begin{eqnarray}
f_{B_s}F_{ann}^{LL,T}(a_i)&=&-f_{B_s}F_{ann}^{LR,T}(a_i)
\end{eqnarray}

\begin{eqnarray}
f_{B_s}F_{ann}^{LL,T}(a_i)
&=& -16\pi C_FM_{B_s}^4f_{B_s}r_2
r_3\int^1_0dx_2dx_3\int^\infty_0b_2db_2b_3db_3 \Big\{
[x_3(\phi_2^v(x_2)\phi_3^v(x_3)+\phi_2^a(x_2)\phi_3^a(x_3))\nonumber\\
&&+(2-x_3)(\phi_2^v(x_2)\phi_3^a(x_3)+\phi_2^a(x_2)\phi_3^v(x_3))]
E_a(t_c)a_i(t_c)h_a(x_2,1-x_3,b_2,b_3)\nonumber\\
&&+h_a(1-x_3,x_2,b_3,b_2)[(1-x_2)
(\phi_2^v(x_2)\phi_3^v(x_3)+\phi_2^a(x_2)\phi_3^a(x_3))
\nonumber\\
&&
-(1+x_2)(\phi_2^v(x_2)\phi_3^a(x_3)+\phi_2^a(x_2)\phi_3^v(x_3))]
E_a(t_c')a_i(t_c')\Big\}.
 \end{eqnarray}

\begin{eqnarray}
f_{B_s} F_{ann}^{LL,L}( a_i)&=&8\pi
C_FM_{B_s}^4f_{B_s}\int^1_0dx_2dx_3\int^\infty_0b_2db_2b_3db_3\Big\{a_i(t_c)
E_a(t_c)
\nonumber\\
&&
\times\Big[(x_3-1)\phi_2(x_2)\phi_3(x_3)-4r_2r_3\phi_2^s(x_2)\phi_3^s(x_3)\nonumber
\\
&&+2r_2r_3x_3\phi_2^s(x_2)(\phi_3^s(x_3)-\phi_3^t(x_3))\Big]h_a(x_2,1-x_3,b_2,b_3)\nonumber
\\
&&+\Big[x_2\phi_2(x_2)
\phi_3(x_3)+2r_2r_3(\phi_2^s(x_2)-\phi_2^t(x_2))\phi_3^s(x_3)\nonumber\\
&&+2r_2r_3x_2(\phi_2^s(x_2)+\phi_2^t(x_2))\phi_3^s(x_3)\Big]
a_i(t_c^\prime)
E_a(t_c^\prime)h_a(1-x_3,x_2,b_3,b_2)\Big\}.\label{ppafll}
 \end{eqnarray}

 \begin{eqnarray}
F_{ann}^{LR,L}( a_i)=F_{ann}^{LL,L}(a_i),\label{ppaflr}
\end{eqnarray}
with the color factor ${C_F} = 3/4$, $f_{B_s}$ refer to the decay constant of $\bar{B}_{s}$ meson and $a_i$ represents the corresponding Wilson coefficients for annihilation decay
 channels. In the above functions, $r_{2} (r_{3}) = m_{V} /m_{B_s}$ and $\phi_{2} (\phi_{3}) = \phi_{V}$ $(V=\rho$ or $\omega)$,
where $m_{V}$ is the chiral scale parameter.

The non-factorizable amplitudes $M_{ann}^{LL,i}(a_i)$, and $M_{ann}^{SP,i}(a_i)$ (i=L,N,T) are written as

\begin{eqnarray}
M_{ann}^{LL,N}(a_i)&=&M_{ann}^{SP,N}(a_i)
\end{eqnarray}

 \begin{eqnarray}
M_{ann}^{LL,N}(a_i)
&=&-64\pi C_FM_{B_s}^4r_2 r_3/\sqrt
{6}\int^1_0dx_1dx_2dx_3\int^\infty_0b_1db_2b_2db_2\phi_{B_s}(x_1,b_1)[\phi_2^v(x_2)\phi_3^v(x_3)\nonumber\\
&&\;\;+\phi_2^a(x_2)\phi_3^a(x_3)]
E_a'(t_d)a_i(t_d)h_{na}(x_1,x_2,x_3,b_1,b_2),
\end{eqnarray}

\begin{eqnarray}
M_{ann}^{LL,T}(a_i)&=& -M_{ann}^{SP,T}(a_i)
\end{eqnarray}

\begin{eqnarray}
M_{ann}^{LL,T}(a_i)
&=&-128 \pi C_FM_{B_s}^4r_2 r_3/\sqrt
{6}\int^1_0dx_1dx_2dx_3\int^\infty_0b_1db_2b_2db_2\phi_{B_s}(x_1,b_1)[\phi_2^v(x_2)\phi_3^a(x_3)\nonumber\\
&&\;\;+\phi_2^a(x_2)\phi_3^v(x_3)]
E_a'(t_d)a_i(t_d)h_{na}(x_1,x_2,x_3,b_1,b_2),
\end{eqnarray}

\begin{eqnarray}
 M_{ann}^{LL,L}( a_i)&=&32\pi C_FM_{B_s}^4/\sqrt
 {6}\int^1_0dx_1dx_2dx_3\int^\infty_0b_1db_2b_2db_2\phi_{B_s}(x_1,b_1)\nonumber\\
 &&\times \Big\{h_{na}(x_1,x_2,x_3,b_1,b_2)\Big[-x_2\phi_2(x_2)\phi_3(x_3)-4r_2r_3
 \phi_2^s(x_2)\phi_3^s(x_3)\nonumber\\
 &&\;\;\;+r_2r_3(1-x_2)(\phi_2^s(x_2)+\phi_2^t(x_2))(\phi_3^s(x_3)-\phi_3^t(x_3))
 \nonumber\\
 &&\;\;+r_2r_3x_3(\phi_2^s(x_2)-\phi_2^t(x_2))(\phi_3^s(x_3)+\phi_3^t(x_3))\Big]a_i(t_d)
 E_a^\prime(t_d)\nonumber\\
 &&\;\;+h_{na}^\prime(x_1,x_2,x_3,b_1,b_2)\Big[(1-x_3)\phi_2(x_2)\phi_3(x_3)
 \nonumber\\
 &&\;\;+(1-x_3)r_2r_3(\phi_2^s(x_2)+\phi_2^t(x_2))(\phi_3^s(x_3)-\phi_3^t(x_3))
 \nonumber\\
 &&\;\;+x_2r_2r_3(\phi_2^s(x_2)-\phi_2^t(x_2))(\phi_3^s(x_3)+\phi_3^t(x_3))\Big]
 a_i(t_d^\prime)
 E_a^\prime(t_d^\prime)\Big\},\label{mlllann}
 \end{eqnarray}

 \begin{eqnarray}
 M_{ann}^{SP,L}( a_i)&=&32\pi C_F M_{B_s}^4/\sqrt {6}\int^1_0dx_1dx_2dx_3\int^\infty_0b_1db_1b_2db_2
 \phi_{B_s}(x_1,b_1)\nonumber\\
 &&\times \Big\{a_i(t_d)E_a^\prime(t_d)h_{na}(x_1,x_2,x_3,b_1,b_2)\Big[(x_3-1)
 \phi_2(x_2)\phi_3(x_3)\nonumber\\
 &&\;\; -4r_2r_3\phi_2^s(x_2)\phi_3^s(x_3)+r_2r_3x_3(\phi_2^s(x_2)+\phi_2^t(x_2))
 (\phi^s_3(x_3)-\phi_3^t(x_3))\nonumber\\
 &&\;\;+r_2r_3(1-x_2)(\phi_2^s(x_2)-\phi_2^t(x_2))(\phi^s_3(x_3)+\phi_3^t(x_3))\Big]
 \nonumber\\
 &&\;\;+a_i(t_d^\prime)
 E_a^\prime(t_d^\prime)h_{na}^\prime(x_1,x_2,x_3,b_1,b_2)
  \Big[x_2\phi_2(x_2)\phi_3(x_3)\nonumber
 \\
 &&\;\;+x_2r_2r_3(\phi_2^s(x_2)+\phi_2^t(x_2))
 (\phi_3^s(x_3)-\phi_3^t(x_3)))\nonumber\\
 &&\;\;+r_2r_3(1-x_3)(\phi_2^s(x_2)-\phi_2^t(x_2))(\phi_3^s(x_3)+\phi_3^t(x_3))\Big]\Big\}.
 \label{ppansp}
 \end{eqnarray}

The hard scale t are chosen as the maximum of the virtuality of the internal momentum transition in the hard amplitudes, including $1/b_i$:
\begin{eqnarray}
t_a&=&\mbox{max}\{{\sqrt
{x_3}M_{B_s},1/b_1,1/b_3}\},\\
t_a^\prime&=&\mbox{max}\{{\sqrt
{x_1}M_{B_s},1/b_1,1/b_3}\},\\
t_b&=&\mbox{max}\{\sqrt
{x_1x_3}M_{B_s},\sqrt{|1-x_1-x_2|x_3}M_{B_s},1/b_1,1/b_2\},\\
t_b^\prime&=&\mbox{max}\{\sqrt{x_1x_3}M_{B_s},\sqrt
{|x_1-x_2|x_3}M_{B_s},1/b_1,1/b_2\},\\
t_c&=&\mbox{max}\{\sqrt{1-x_3}M_{B_s},1/b_2,1/b_3\},\\
t_c^\prime
&=&\mbox{max}\{\sqrt {x_2}M_{B_s},1/b_2,1/b_3\},\\
t_d&=&\mbox{max}\{\sqrt {x_2(1-x_3)}M_{B_s},
\sqrt{1-(1-x_1-x_2)x_3}M_{B_s},1/b_1,1/b_2\},\\
t_d^\prime&=&\mbox{max}\{\sqrt{x_2(1-x_3)}M_{B_s},\sqrt{|x_1-x_2|(1-x_3)}M_{B_s},1/b_1,1/b_2\}.
\end{eqnarray}

The hard functions $h$ are written as \cite{L3}
\begin{eqnarray}
h_e(x_1,x_3,b_1,b_3)&=&\left[\theta(b_1-b_3)I_0(\sqrt
x_3M_{B_s}b_3)K_0(\sqrt
x_3 M_{B_s}b_1)\right.\\
&& \left.+\theta(b_3-b_1)I_0(\sqrt x_3M_{B_s}b_1)K_0(\sqrt
x_3M_{B_s}b_3)\right]K_0(\sqrt {x_1x_3}M_{B_s}b_1)S_t(x_3),\nonumber\\
h_n(x_1,x_2,x_3,b_1,b_2)&=&\left[\theta(b_2-b_1)K_0(\sqrt
{x_1x_3}M_{B_s}b_2)I_0(\sqrt
{x_1x_3}M_{B_s}b_1)\right. \nonumber\\
&&\;\;\;\left. +\theta(b_1-b_2)K_0(\sqrt
{x_1x_3}M_{B_s}b_1)I_0(\sqrt{x_1x_3}M_{B_s}b_2)\right]\nonumber\\
&&\times
\left\{\begin{array}{ll}\frac{i\pi}{2}H_0^{(1)}(\sqrt{(x_2-x_1)x_3}
M_{B_s}b_2),& x_1-x_2<0\\
K_0(\sqrt{(x_1-x_2)x_3}M_{B_s}b_2),& x_1-x_2>0
\end{array}
\right. ,
\end{eqnarray}
\begin{eqnarray}
h_a(x_2,x_3,b_2,b_3)&=&(\frac{i\pi}{2})^2
S_t(x_3)\Big[\theta(b_2-b_3)H_0^{(1)}(\sqrt{x_3}M_{B_s}b_2)J_0(\sqrt
{x_3}M_{B_s}b_3)\nonumber\\
&&\;\;+\theta(b_3-b_2)H_0^{(1)}(\sqrt {x_3}M_{B_s}b_3)J_0(\sqrt
{x_3}M_{B_s}b_2)\Big]H_0^{(1)}(\sqrt{x_2x_3}M_{B_s}b_2),\\
h_{na}(x_1,x_2,x_3,b_1,b_2)&=&\frac{i\pi}{2}\left[\theta(b_1-b_2)H^{(1)}_0(\sqrt
{x_2(1-x_3)}M_{B_s}b_1)J_0(\sqrt {x_2(1-x_3)}M_{B_s}b_2)\right. \nonumber\\
&&\;\;\left.
+\theta(b_2-b_1)H^{(1)}_0(\sqrt{x_2(1-x_3)}M_{B_s}b_2)J_0(\sqrt
{x_2(1-x_3)}M_{B_s}b_1)\right]\nonumber\\
&&\;\;\;\times K_0(\sqrt{1-(1-x_1-x_2)x_3}M_{B_s}b_1),\\
h_{na}^\prime(x_1,x_2,x_3,b_1,b_2)&=&\frac{i\pi}{2}\left[\theta(b_1-b_2)H^{(1)}_0(\sqrt
{x_2(1-x_3)}M_{B_s}b_1)J_0(\sqrt{x_2(1-x_3)}M_{B_s}b_2)\right. \nonumber\\
&&\;\;\;\left. +\theta(b_2-b_1)H^{(1)}_0(\sqrt
{x_2(1-x_3)}M_{B_s}b_2)J_0(\sqrt{x_2(1-x_3)}M_{B_s}b_1)\right]\nonumber\\
&&\;\;\;\times
\left\{\begin{array}{ll}\frac{i\pi}{2}H^{(1)}_0(\sqrt{(x_2-x_1)(1-x_3)}M_{B_s}b_1),&
x_1-x_2<0\\
K_0(\sqrt {(x_1-x_2)(1-x_3)}M_{B_s}b_1),&
x_1-x_2>0\end{array}\right. ,
\end{eqnarray}
where $J_0$ and ${Y}_0$ are the Bessel function with $H_0^{(1)}(z) = \mathrm{J}_0(z) + i\, \mathrm{Y}_0(z)$.

The threshold re-sums factor $S_t$ follows the parameterized \cite{Kurimoto:2001zj}
\begin{eqnarray}
S_t(x)=\frac{2^{1+2c}\Gamma(3/2+c)}{\sqrt \pi \Gamma(1+c)}[x(1-x)]^c,
\end{eqnarray}
where the parameter $c=0.4$. In the nonfactorizable contributions, $S_t(x)$ gives a very small numerical effect to the amplitude~\cite{L4}.
Therefore, we drop $S_t(x)$ in $h_n$ and $h_{na}$.

The evolution factors $E^{(\prime)}_e$ and $E^{(\prime)}_a$ entering in the expressions for the matrix elements are given by
\begin{eqnarray}
E_e(t)&=&\alpha_s(t) \exp[-S_B(t)-S_3(t)], \ \ \ \ E'_e(t)=\alpha_s(t) \exp[-S_B(t)-S_2(t)-S_3(t)]|_{b_1=b_3},\\
E_a(t)&=&\alpha_s(t) \exp[-S_2(t)-S_3(t)],\ \ \ \ E'_a(t)=\alpha_s(t) \exp[-S_B(t)-S_2(t)-S_3(t)]|_{b_2=b_3},
\end{eqnarray}
in which the Sudakov exponents are defined as
\begin{eqnarray}
S_B(t)&=&s\left(x_1\frac{M_{B_s}}{\sqrt2},b_1\right)+\frac{5}{3}\int^t_{1/b_1}\frac{d\bar \mu}{\bar\mu}\gamma_q(\alpha_s(\bar \mu)),\\
S_2(t)&=&s\left(x_2\frac{M_{B_s}}{\sqrt2},b_2\right)+s\left((1-x_2)\frac{M_{B_s}}{\sqrt2},b_2\right)+2\int^t_{1/b_2}\frac{d\bar \mu}{\bar
\mu}\gamma_q(\alpha_s(\bar \mu)),
\end{eqnarray}
where $\gamma_q=-\alpha_s/\pi$ is the anomalous dimension of the quark. The explicit form for the  function $s(Q,b)$ is:
\begin{eqnarray}
s(Q,b)&=&\frac{A^{(1)}}{2\beta_{1}}\hat{q}\ln\left(\frac{\hat{q}}{\hat{b}}\right)-\frac{A^{(1)}}{2\beta_{1}}\left(\hat{q}-\hat{b}\right)+
\frac{A^{(2)}}{4\beta_{1}^{2}}\left(\frac{\hat{q}}{\hat{b}}-1\right)
-\left[\frac{A^{(2)}}{4\beta_{1}^{2}}-\frac{A^{(1)}}{4\beta_{1}}
\ln\left(\frac{e^{2\gamma_E-1}}{2}\right)\right]
\ln\left(\frac{\hat{q}}{\hat{b}}\right)
\nonumber \\
&&+\frac{A^{(1)}\beta_{2}}{4\beta_{1}^{3}}\hat{q}\left[
\frac{\ln(2\hat{q})+1}{\hat{q}}-\frac{\ln(2\hat{b})+1}{\hat{b}}\right]
+\frac{A^{(1)}\beta_{2}}{8\beta_{1}^{3}}\left[
\ln^{2}(2\hat{q})-\ln^{2}(2\hat{b})\right],
\end{eqnarray}
where the variables are defined by
\begin{eqnarray}
\hat q\equiv \mbox{ln}[Q/(\sqrt 2\Lambda)],~~~ \hat b\equiv
\mbox{ln}[1/(b\Lambda)], \end{eqnarray} and the coefficients
$A^{(i)}$ and $\beta_i$ are \begin{eqnarray}
\beta_1=\frac{33-2n_f}{12},~~\beta_2=\frac{153-19n_f}{24},\nonumber\\
A^{(1)}=\frac{4}{3},~~A^{(2)}=\frac{67}{9}
-\frac{\pi^2}{3}-\frac{10}{27}n_f+\frac{8}{3}\beta_1\mbox{ln}(\frac{1}{2}e^{\gamma_E}),
\end{eqnarray}
with $n_f$ is the number of the quark flavors and $\gamma_E$ is the
Euler constant. We will use the one-loop expression of the running coupling constant.

In this study, we use the model function
\begin{equation}
\phi_{B_s}(x,b) = N_{B_s} x^2(1-x)^2 \exp \left[ -\frac{M_{B_s}^2\
x^2}{2 \omega_b^2} -\frac{1}{2} (\omega_b b)^2 \right],\label{waveb}
\end{equation}
where the share parameter $\omega_b=0.5\pm 0.05$ GeV, and the normalization constant $N_{B_s}=63.5688$ GeV is related to the $B_s$ decay constant ${f_{{B_s}}} = 0.23 \pm 0.03$ GeV.

For $\rho$ and $\omega$ vector meson, we use ${\rho ^0} = \frac{1}{{\sqrt 2 }}\left( {u\overline u  - d\overline d } \right)$ and $\omega  = \frac{1}{{\sqrt 2 }}\left( {u\overline u  + d\overline d } \right)$. The distribution amplitudes of vector meson(v=$\rho$ or $\omega$), $\phi_{\rho}$, $\phi_{\omega}$, $\phi^t_{V}$, $\phi^s_{V}$, $\phi^v_{V}$, and $\phi^a_{V}$,
are calculated using light-cone QCD sum rule \cite{hep-ph/0412079}:
\begin{eqnarray}
\phi_\rho (x)&=&\frac{3f_\rho}{\sqrt{6}} x (1-x)\left[1+0.15C_2^{3/2} (t) \right]\;,\label{phirho}\\
\phi_\omega(x)&=&\frac{3f_\omega}{\sqrt{6}} x (1-x)\left[1+0.15C_2^{3/2} (t)\right]\;,\label{phiomega}\\
\phi^t_V(x)&=&\frac{3f^T_V}{2\sqrt 6}t^2\;,\label{phitv}\\
\phi^s_V(x)&=&\frac{3f_V^T}{2\sqrt 6} (-t)\;,\label{phisv}\\
\phi_V^v(x)&=&\frac{3f_V}{8\sqrt6}(1+t^2)\;,\label{phivv}\\
 \phi_V^a(x)&=&\frac{3f_V}{4\sqrt6}(-t)\;,\label{phiav}
\end{eqnarray}
where $t=2x-1$. Here $f_{V}$ is the decay constant of
the vector meson with longitudinal  polarization, whose values are
shown in table \ref{table1}.

The Gegenbauer polynomials $C^{\nu}_{n}(t)$ read,
\begin{equation}
 \begin{array}{ll}
 C_2^{1/2} (t) = \frac{1}{2} (3t^2-1)\;,\;\;\;& C_4^{1/2} (t) = \frac{1}{8}(35t^4-30t^2+3),\\
 C_2^{3/2} (t) = \frac{3}{2} (5t^2-1)\;,\;\;\;& C_4^{3/2} (t) = \frac{15}{8}(1-14t^2+21t^4),\\
 C_1^{3/2}(t)= 3t.
 \end{array}
 \end{equation}

\begin{acknowledgments}
This work was supported by National Natural Science
Foundation of China (Project Numbers 11605041), Plan For Scientific Innovation Talent of Henan University of Technology
 (Project Number 2012CXRC17), the Key Project (Project Number 14A140001)
 for Science and Technology of the Education Department Henan Province,
the Fundamental Research Funds (Project Number 2014YWQN06) for the Henan Provincial Colleges and Universities,
and the Research Foundation of the young core teacher from Henan province.
\end{acknowledgments}


\end{spacing}
\end{document}